\magnification=1200
\def\qed{\unskip\kern 6pt\penalty 500\raise -2pt\hbox
{\vrule\vbox to 10pt{\hrule width 4pt\vfill\hrule}\vrule}}
\centerline{A REMARK ON THE EQUIVALENCE OF ISOKINETIC AND ISOENERGETIC}
\centerline{THERMOSTATS IN THE THERMODYNAMIC LIMIT.}
\bigskip
\centerline{by David Ruelle\footnote{*}{IHES.  91440 Bures sur Yvette,
France. $<$ruelle@ihes.fr$>$}.}
\bigskip\bigskip\noindent
	{\leftskip=2cm\rightskip=2cm\sl Abstract.  The Gaussian isokinetic and isoenergetic thermostats of Hoover and Evans are formally equivalent as remarked by Gallavotti, Rondoni and Cohen.  But outside of equilibrium the fluctuations are uncontrolled and might break the equivalence.  We show that equivalence is ensured if we consider an infinite system assumed to be ergodic under space translations.\par}
\vfill\eject
\bigskip\bigskip
	{\bf 1. Introduction.}
\medskip
	In the study of nonequilibrium statistical mechanics, if
nonhamiltonian forces are used to achieve nonequilibrium, a thermostat
is needed to cool the system.  The {\it Gaussian thermostats} introduced
by W. Hoover and D. Evans have the great interest of respecting the
deterministic character of the equations of motion (see for instance
Evans and Morriss [3]).  Starting with an evolution equation $\dot
x=F(x)$ in phase space, a Gaussian thermostat constrains the evolution
to a prescribed hypersurface $\Sigma$ by projecting $F(x)$, for
$x\in\Sigma$, to the tangent plane to $\Sigma$ at $x$.  In the present
note we follow Cohen-Rondoni, and Gallavotti comparing an isokinetic and an isoenergetic thermostat, and showing that they give the same result in the limit of a large system (thermodynamic limit).
\medskip
	In equilibrium statistical mechanics one can show rigorously
that fixing the kinetic energy is equivalent to fixing the total energy,
asymptotically for large systems (see [6]).  It is therefore natural to
hope that something similar is true for nonequilibrium, as advocated by
Gallavotti (many references, see [4], [5]) and by Cohen and Rondoni [2].
However, the entropy considerations which are available in equilibrium 
statistical mechanics fail utterly outside of equilibrium, {\it i.e.}, 
fluctuations of energy at fixed kinetic energy are uncontrolled, and the
situation appears rather hopeless.  We shall show however that the
argument of Cohen and Rondoni can be modified to apply, at least formally, to the dynamics of actually infinite systems.  (In a different context -- at equilibrium -- Sinai [7] has also shown the interest of considering the dynamics of infinite systems).  Our approach will remain formal at the level of infinite system evolution equations: technical problems arise there, which do not seem directly related to the problem at hand, and are better discussed separately.
\medskip
	We shall consider a system of particles in $d$ dimensions which is infinitely extended in $\nu$ dimensions, with $1\le\nu\le d$, and we shall discuss states of infinitely many particles which are invariant under translations in ${\bf R}^\nu$.  The assumption that the infinite systems dynamics is well defined, and ${\bf R}^\nu$-ergodicity, will be sufficient to establish the equivalence of isokinetic and isoenergetic nonequilibrium steady states
\medskip
	{\bf 2. IK an IE dynamics.}
\medskip
	We recall now the definition of the {\it Gaussian isokinetic}
(IK) thermostat.  We take for our configuration space $M$ a compact
subset of ${\bf R}^u\times{\bf T}^v$ where ${\bf T}^v$ is the $v$-torus,
and momentum space is identified with ${\bf R}^{u+v}$.  We assume that a
force field on $M$ is given, written as $-{\rm grad}V+\xi$, where
$V:M\to{\bf R}$ is a potential, and $\xi$ is a nongradient vector
field\footnote{*}{Note that a change in $V$ can be compensated by a
corresponding change in $\xi$: the splitting of the force into two terms
is arbitrary for the IK time evolution.}.  Consider now the equations of motion
$$	\matrix{\dot p=-\partial_qV+\xi-\alpha p\cr
	\dot q=p/m}\eqno{(1)}      $$
completed by elastic reflection at the boundary of $M$.  Without the
term $\xi-\alpha p$ this time evolution would be Hamiltonian.  The term
$\xi$ maintains the system outside of equilibrium.  The term $-\alpha p$
is the thermostat.  We obtain the Gausssian isokinetic thermostat by
choosing $\alpha$ such that the kinetic energy is constant:
$$	0={d\over dt}{p^2\over2m}
	={p\over m}\cdot(-\partial_qV+\xi-\alpha p)      $$
{\it i.e.},
$$	\alpha=(-\partial_qV+\xi)\cdot p/p^2\eqno{(2)}      $$
Note that if $\xi$ is locally a gradient (corresponding to a multivalued potential function on $M$), the Dettmann-Morriss pairing theorem asserts that (except for one value $=0$) the spectrum of Lyapunov exponents of an ergodic measure is symmetric with respect to some constant $c$ which is in general nonzero.  (We shall however not make use of this result).
\medskip
	We consider now the {\it Gaussian isoenergetic} (IE) thermostat
associated again with the force $-\hbox{grad}V+\xi$, but where we want to
maintain fixed the energy function
$$	H=p^2/2m+V(q)\eqno{(3)}      $$
The equations of motion are again of the form (1) and using (3) the isoenergetic condition is 
$$	0=\dot H={p\over m}\cdot(-\partial_qV+\xi-\alpha p)
	+\partial_qV\cdot{p\over m}      $$
{\it i.e.},
$$	\alpha=\xi\cdot p/p^2\eqno{(4)}      $$
With the Gaussian isoenergetic (IE) thermostat the time evolution is thus defined by (1), (4). 
\medskip
	We consider now the IK and the IE time evolution in the infinite
system limit.  We want to study the time evolution of a state $\rho$
ergodic under ${\bf R}^\nu$-space translations.  We shall ignore
existence and uniqueness problems for these evolution equations, and our
discussion will thus remain formal in this respect.  (In fact, the
one-dimensional situation may be relatively accessible to rigorous
study, but the $n$-dimensional case with $n\ge2$ appears much more difficult).
\medskip
	Physically we may think of a system of particles in a region $D$
invariant under ${\bf R}^\nu$, where $1\le\nu\le\hbox{dim}D$ but
possibly $\nu<\hbox{dim}D$.  For example we may consider a shear flow
between two moving plates, but we do not take the limit where these two
plates are infinitely far apart, as this would introduce unwanted
hydrodynamic instabilities.  Another example would be a system of
particles in $[0,L]\times{\bf R}^\nu$.  In the $x$-direction we put an
electric field and we assume a suitable boundary condition (see [1]).
\medskip
	The expressions $p^2=p\cdot p$ and $\xi\cdot p$ diverge for an infinite system, but behave additively with respect to volume, and we can (under mild conditions on $\rho$) define the average per unit volume with respect to $\rho$, noted $\langle p^2\rangle_\rho$ or $\langle\xi\cdot p\rangle_\rho$.  Since $\rho$ is ergodic, it is carried by points (in infinite phase space) for which the large volume average of $p^2$ or $\xi\cdot p$ is well defined and constant, equal to $\langle p^2\rangle_\rho$ or $\langle\xi\cdot p\rangle_\rho$.  The expressions $V$, $\partial_qV\cdot p$ behave almost additively with respect to volume and, again under mild conditions, we can define the large volume averages $\langle V\rangle_\rho$, $\langle\partial_qV\cdot p\rangle_\rho$.  Again, $\rho$ is carried by points (in infinite phase space) for which the large volume average of $V$ or $\partial_qV\cdot p$ is well defined and constant, equal to $\langle V\rangle_\rho$ or $\langle\partial_qV\cdot p\rangle_\rho$.
\medskip
	In our formal treatment of the infinite system IK or IE evolution we consider the time evolution of an infinite phase space point, generic with respect to the space ergodic measure $\rho$, replacing the expressions (2), (4) for $\alpha$ by their large volume limits
$$	\alpha=\langle(-\partial_qV+\xi)\cdot p\rangle_\rho
	/\langle p^2\rangle_\rho\eqno{(2')}      $$
or
$$	\alpha=\langle\xi\cdot p\rangle_\rho
	/\langle p^2\rangle_\rho\eqno{(4')}      $$
In general $\rho$ depends on time, and so does $\alpha$ given by $(2')$ or $(4')$.  Suppose now that $\rho$ is invariant under the IK or IE time evolution; then $\alpha$ and also $V$ are time independent, so that 
$$	0=\langle\dot V\rangle_\rho=\langle\partial_qV\cdot\dot q\rangle_\rho
	={1\over m}\langle\partial_qV\cdot p\rangle_\rho      $$
But then $(2')$ and $(4')$ coincide: the infinite system IK and IE
evolutions have the same time invariant space ergodic states $\rho$.
(Apart from the use of space ergodicity for an actually infinite system,
this is the remark of Cohen and Rondoni [2]).
\medskip
	Note that, if we replace in (3) $m$ by $\tilde m$ and $V$ by
$\tilde V$, imposing $\dot H=0$ yields 
$$	\alpha=({\tilde m\over m}
	\partial_q\tilde V-\partial_q V+\xi)\cdot{p\over p^2}      $$
and in the infinite system limit we have again equivalence with the
isokinetic ensemble.  On the other hand, if $H$ is not of the form
$p^2/2\tilde m+\tilde V(q)$, the Gaussian thermostat doesn't give a term
of the form $-\alpha p$ in (1) and we do not have equivalence with the
isokinetic ensemble in the infinite system limit. 
\medskip
	For the purposes of nonequilibrium statistical mechanics one should presumably restrict $\rho$ to be an infinite system SRB state (defined so that the time entropy per unit volume is equal to the sum of the positive Lyapunov exponents per unit volume).  Hopefully, the space ergodic SRB states form a 2-parameter family parametrized by the average number of particles and the energy (or the kinetic energy) per unit volume.  But the delicate question of identifying the natural nonequilibrium steady states is here bypassed by the remark that they are the same for the infinite system IK and IE evolutions.
\medskip
	In equilibrium statistical mechanics the proof of equivalence of ensembles is somewhat subtle, and uses in particular the concavity properties of the entropy (see [6]).  One might think that the corresponding problem in nonequilibrium statistical mechanics would be even more difficult, and the above findings about the equivalence of IK and IE appear thus surprisingly cheap.  What we have shown is however only that the IK and IE evolutions coincide (formally) in the infinite system limit; the detailed study of the natural nonequilibrium states remains to be made.
\medskip
	{\bf 3. The constant $\alpha$ case.}
\medskip
	It is of interest to consider the equations (1) with $\alpha$ = constant.  For this situation one obtains the following result.
\medskip
	{\bf Proposition.}
\medskip
	{\sl Consider the evolution equations
$$	\matrix{\dot p=-\partial_qV+\xi-\alpha p\cr\dot q=p/m}      $$
in $TM$, where $M\subset{\bf R}^u\times{\bf T}^v$ and we impose elastic reflection on the boundary of $M$.  We assume that $\alpha$, $m$ are constants $>0$, and that $V$, $\xi$ are bounded.  Then
$$	{\lim\sup}_{t\to\infty}({p^2\over2m}+V)\le
	{\max\xi^2\over2m\alpha^2}+\max V\eqno{(5)}      $$
$$	{\lim\sup}_{t\to\infty}p^2\le
	{\max\xi^2\over\alpha^2}+2m(\max V-\min V)\eqno{(6)}      $$
Furthermore, if the bounded measure $\rho$ is invariant under time evolution and, and $\Phi$ is any continuous function we have
$$	\int\rho(dp\,dq)\Phi({p^2\over2m}).(\xi\cdot p-\alpha p^2)
	=0\eqno{(7)}      $$}
\medskip
	From the evolution equations we obtain
$$	{d\over dt}({p^2\over2m}+V)
	={p\over m}\cdot\dot p+\partial_q\cdot\dot q
	={p\over m}\cdot(-\partial_qV+\xi-\alpha p)+\partial_q\cdot{p\over m}
	={p\over m}\cdot(\xi-\alpha p)\eqno{(8)}      $$
Let now $\epsilon>0$ and suppose that 
$$  {p^2\over2m}+V\ge{\max\xi^2\over2m\alpha^2}+\max V+\epsilon\eqno{(9)}  $$
then
$$	p^2\ge\max\xi^2/\alpha^2+\epsilon      $$
or
$$	\alpha|p|\ge\max|\xi|+\epsilon'      $$
with $\epsilon'>0$ and thus, in view of (8),
$$	{d\over dt}({p^2\over2m}+V)\le{1\over m}(|p||\xi|-\alpha|p|^2)
	\le-{|p|\over m}\,\epsilon'      $$
Therefore, as long as (9) holds, we have
$$	{d\over dt}({p^2\over2m}+V)\le-\delta      $$
for some $\delta>0$, proving (5).  From (5) we obtain immediately (6).
\medskip
	Let $\Psi'=\Phi$ then, by the ergodic theorem,
$$	\int\rho(dp\,dq)\Phi({p^2\over2m}).(\xi\cdot p-\alpha p^2)
	=\lim_{T\to\infty}{1\over T}\int_0^Tdt\,\Phi({p^2\over2m})
	.(\xi\cdot p-\alpha p^2)      $$
$$	=\lim_{T\to\infty}{m\over T}\int_0^Tdt\,\Psi'({p^2\over2m})
	.{d\over dt}({p^2\over2m}+V)
	=\lim_{T\to\infty}{m\over T}\int_0^Tdt\,
	{d\over dt}\Psi({p^2\over2m}+V)      $$
$$	=\lim_{T\to\infty}{m\over T}[\Psi({p^2\over2m}+V)]_0^\infty=0      $$
because $\Psi$ is bounded in view of (5).  This proves (7).\qed
\medskip
	{\bf Acknowledgements.}
\medskip
	I am indebted to Eddie Cohen, Giovanni Gallavotti, and Oscar
Lanford for enlightening discussions concerning this note.
\medskip
	{\bf References.}
[1] N.I. Chernov, G.L. Eyink, J.L. Lebowitz, and Ya.G. Sinai.  ``Derivation of Ohms law in a deterministic mechanical model.''  Phys. Rev. Letters {\bf 70}, 2209-2212(1993).

[2] E.G.D. Cohen and L. Rondoni.  ``Note on phase space contraction and entropy production in thermostatted Hamiltonian systems.''  Chaos {\bf 8},357-365(1998).

[3] D.J. Evans and G.P. Morriss.  {\it Statistical mechanics of nonequilibrium fluids.}  Academic Press, New York, 1990.

[4] G. Gallavotti.  ``Dynamical ensembles equivalence in fluid mechanics.''  Physica D {\bf 105},163-184(1997).

[5] G. Gallavotti.  ``Chaotic dynamics, fluctuations, non-equilibrium ensembles.''\break  Chaos {\bf 8},384-392(1998).

[6] D. Ruelle.  ``Correlation functionals.''  J. Math. Phys. {\bf 6},201-220(1965).

[7] Ya.G. Sinai.  ``A remark concerning the thermodynamic limit of the Lyapunov spectrum.''  Int. J. Bifurc. Chaos {\bf 6},1137-1142(1996).
\end